\documentclass[letterpaper,12pt]{article}
\usepackage{osajnl2}
\usepackage[singlelinecheck=false, position=top, caption=false, captionskip=-0.15in]{subfig}
\usepackage[breaklinks=true, draft]{hyperref}

\newcommand{\confA}{configuration~\textit{A}}
\newcommand{\confB}{configuration~\textit{B}}
\newcommand{\confC}{configuration~\textit{C}}

\begin{document}

\title{Dynamically configurable and optimizable Zeeman slower using permanent magnets and servomotors}

\author{G. Reinaudi, C. B. Osborn, K. Bega, and T. Zelevinsky$^*$}

\address{Department of Physics, Columbia University, 538 West 120th Street, New York, NY 10027}
\address{$^*$Corresponding author: tz@phys.columbia.edu}



\begin{abstract}
We report on the implementation of a dynamically configurable, servomotor-controlled, permanent magnet Zeeman slower for quantum optics experiments with ultracold atoms and molecules. This atom slower allows for switching between magnetic field profiles that are designed for different atomic species. Additionally, through feedback on the atom trapping rate, we demonstrate that computer-controlled genetic optimization algorithms applied to the magnet positions can be used in situ to obtain field profiles that maximize the trapping rate for any given experimental conditions. The device is lightweight, remotely controlled, and consumes no power in steady state; it is a step toward automated control of quantum optics experiments.
\end{abstract}


\maketitle


\section{Introduction}
\label{sec:intro}

Laser cooling and trapping of neutral atoms \cite{Phillips98, Wieman99} has led to rapid progress in the fields of quantum optics, precision metrology, and ultracold chemistry.  Many experiments in these fields use collimated beams of atoms that are slowed and trapped in magneto-optical traps (MOTs).  The classic approach to atomic beam slowing uses radiation pressure from counter-propagating, near-resonant laser light, combined with a shaped magnetic field profile that compensates for the changing Doppler shift of the atoms via the Zeeman effect.  This magnetic field shaper, known as a Zeeman slower (ZS) \cite{Bagnato89, DNH04}, is traditionally constructed by winding carefully designed layers of wire around the atomic beam vacuum tube, and running a current through the wire. Recently, designs based on permanent magnets have been suggested and implemented \cite{Ovc08, Guery11}. There are multiple benefits to using a permanent magnet ZS. These benefits span adaptability, robustness, ease of maintenance, zero power consumption, and low cost; the current-sourcing power supply as well as water cooling are not necessary. With the plans to bring state-of-the-art atomic clocks to space in the near future \cite{SOC10}, these attributes become critical.

Many modern quantum optics experiments demand the flexibility of cooling and trapping multiple atomic species.  The applications include creation of ultracold molecules \cite{Carr09} for studies of novel phases and quantum chemical reactions, highly precise multi-species atomic clocks \cite{YudinPRL11}, and calibration of the trapping apparatus for ultrasensitive low-level atom counting \cite{ColumbiaATTA}.  In this work, we present an implementation of a dynamically controlled, permanent magnet ZS designed for slowing multiple atomic species for any of the above applications.  Furthermore, the magnetic field profile produced by the ZS is optimizable in real time, mitigating any experimental imperfections not accounted for in the design. This approach is a step toward building fully automated, self-optimizing, quantum optics experiments.

This paper is organized as follows. Section 2 contains a brief summary of the ZS design. In Sec. 3 we describe the implementation of the magnetic field computer control through the use of servomotors. In Sec. 4 we demonstrate the automatic field optimization via computer feedback.  Section 5 contains the conclusions.

\section{Design considerations}
\label{sec:construction}
\begin{figure}[ht]
\centering
\subfloat[]{
$\begin{array}{c}
\vspace{0.015in} \\
\includegraphics[height=1.8in]{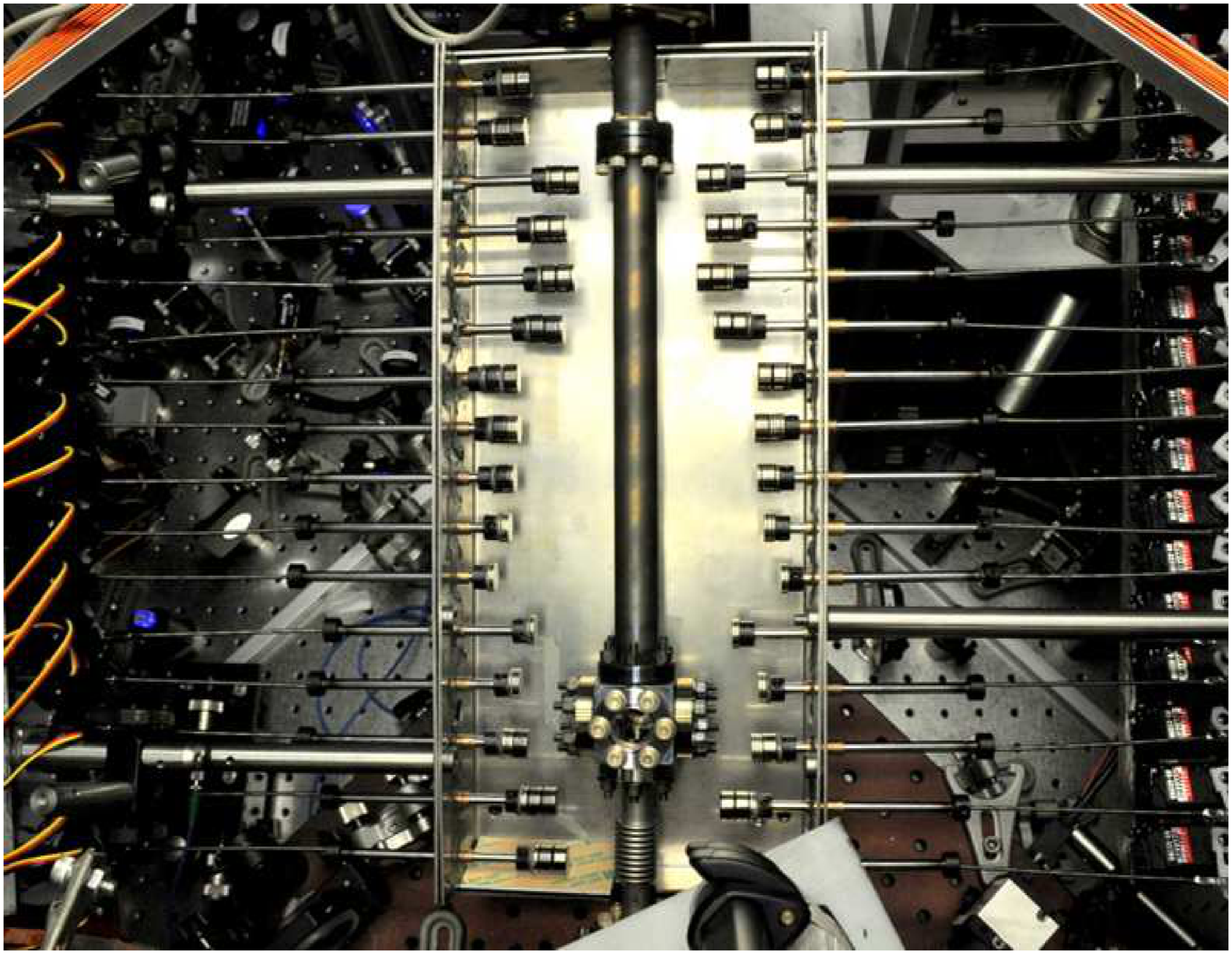}
\end{array}$
\label{fig:ServoPicture}
}
\subfloat[]{
$\begin{array}{c}
\vspace{0.015in} \\
\includegraphics[height=1.8in]{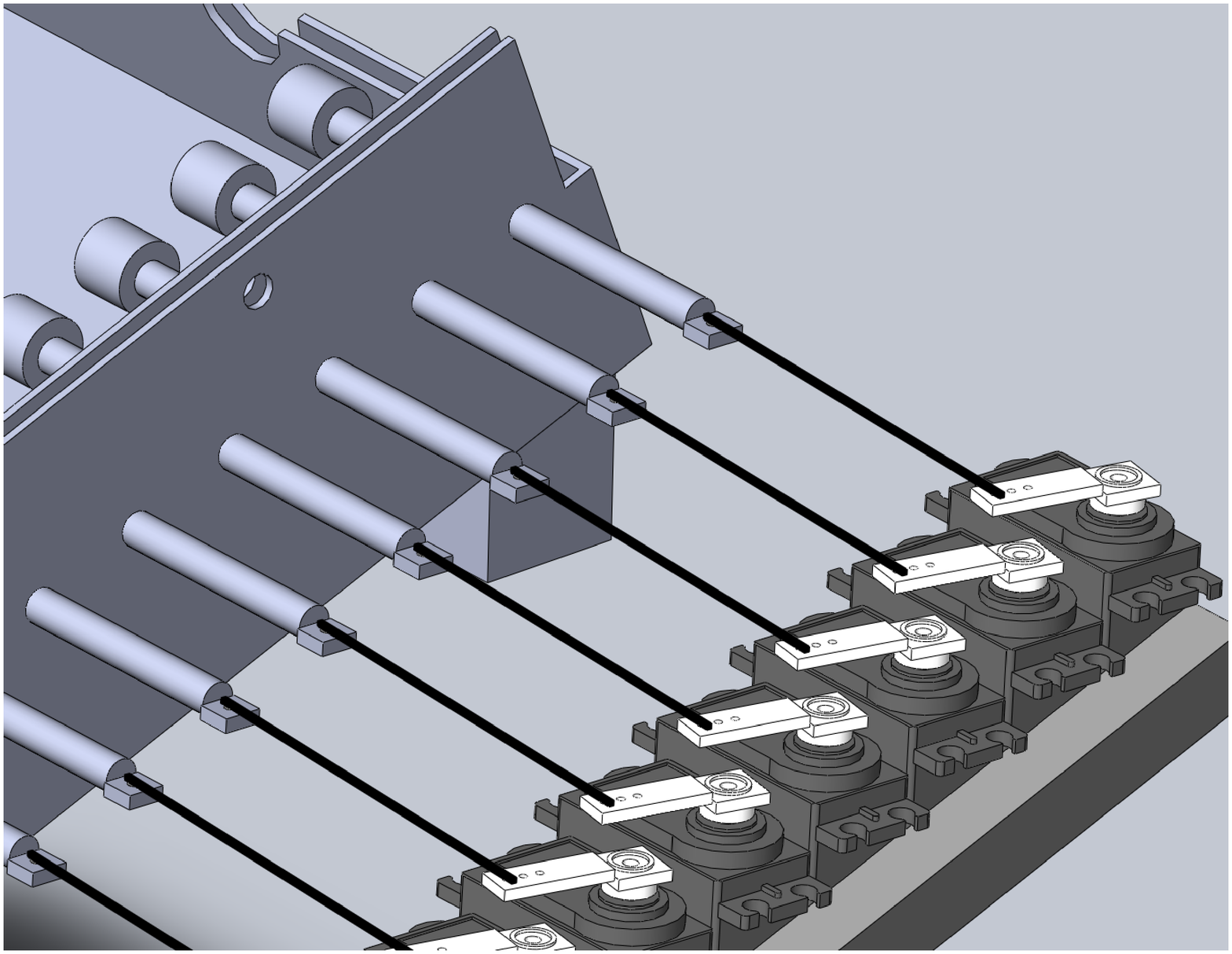}
\end{array}$
\label{fig:ServoViewSolidworks}
}
\caption{
(Color online) 
(a) View of the ZS (lid off), including the 16 pairs of magnets and servomotors surrounding the atomic beam vacuum section. 
(b) Close-up rendering of servomotor actuation.
}
\label{fig:ServosView}
\end{figure}

The magnetic field profile for efficient atom slowing is set by placing an array of permanent magnets at variable distances from the atomic beam. The magnets are neodymium disks of $0.5^{\prime\prime}$ diameter, $1/16$-$1/4^{\prime\prime}$ thickness, and 0.25-0.95 J/T magnetic moment.  The entire assembly is enclosed in a $\mu$-metal magnetic shield, as pictured in Fig. \ref{fig:ServoPicture}.
We performed point-like magnetic dipole computer simulations to determine a ZS design, as described in Ref. \cite{Ovc08}.

Design constraints include the finite strength and size of the magnets, and spatial constraints of the vacuum system. Achieving a smooth field profile as well as a high field strength along the atomic beam path is a balancing act, since the closer a magnet is to the beam axis, the sharper its local effect on the magnetic field and hence the closer neighboring magnets have to be placed for a smoothing effect.
We found that performing the computer simulation is important for determining the right spacing between neighboring magnets, and less important for determining magnet distance from the beam axis, since the latter is adjustable while the former is a fixed feature of the construction.

The magnets were originally fitted onto manually adjustable nylon screws, chosen for their pliability as well as nonmagnetic properties. By manual placement of the screws, a specific magnetic field profile was matched nearly perfectly, with the resultant magnet positions slightly closer to the beam axis than predicted by the computer simulation.

The resulting magnetic field is transverse to the atomic beam. As atoms travel through the field region, they are decelerated by counter-propagating light that is linearly polarized in the direction normal to the field, and gently focused to match the slight divergence of the 6 mm wide atomic beam.

\section{Computer control of the magnetic field profile}
\label{sec:servomotors}

\subsection{Actuation through servomotors}
\label{sec:servomotors-electro}

In order to allow real-time computer control of the magnetic field profile, the magnets were mounted to stainless steel rods that freely slide in bronze sleeve bearings placed through the walls of the $\mu$-metal case (Fig.~\ref{fig:ServoPicture}).
The actuation of the magnet positions relies on low-cost servomotors that are typically used in model airplanes and cars. Such servomotors provide a rotation range of $\pm 90^{\circ}$ and a velocity of $\sim 60^{\circ}$/s. The model we used is HS-422 by Hitec, but there is a large variety of options available depending on different requirements, such as torque, speed and size.

In order to transform the rotary motion of the servomotors into linear motion, each sliding rod is linked to the horn of a servomotor through a clevis and a threaded rod (Fig.~\ref{fig:ServoViewSolidworks}). Sixteen servomotors are placed on each side of the ZS. For ease of assembly, the servomotors are glued to each other (with appropriate spacers) and to an aluminum bracket attached to the case of the ZS.

The servomotors are controlled by electronic modules manufactured by Phidgets. While other options are available, these modules were chosen for their cost effectiveness and ease of use, since they interface with a computer USB port and provide a straightforward application programming interface (API) for most common programming languages. Four modules are needed to control the 32 servomotors.
During the operation of the motors, care must be taken about the range that each of them should be allowed to explore. We limit the movement of the servomotors using the computer program that controls them \cite{LabExe}.

\subsection{Field profile switching for slowing multiple species}
\label{sec:servomotors-switch}
Many modern experiments involve trapping several atomic species. In those cases, the ZS is typically designed for the one that requires a longer slowing region, thus not achieving the maximum efficiency for the other species at this slower length. However, in experiments that routinely change the trapped species it would be useful to switch rapidly between magnetic field profiles in a reproducible way so that each species is optimally slowed.
The computer-controlled setup described here makes this process straightforward since any number of magnet configurations are saved in the software and can be retrieved when needed. Additionally, for experiments that use two species simultaneously, one can sequentially load each species using its optimal profile, since the switching between field profiles is consistently~$<0.5$~s.

In general, designing the ZS magnetic field depends on many practical factors such as the available space and laser power, the initial thermal velocity of the atoms, and the presence of transverse cooling of the atomic beam \cite{DNH04}. Our setup allows for tuning the magnetic profile for an optimal loading given these specific experimental constraints.

\begin{figure}[ht]
\centering
\subfloat[]{
\includegraphics[height=1.65in]{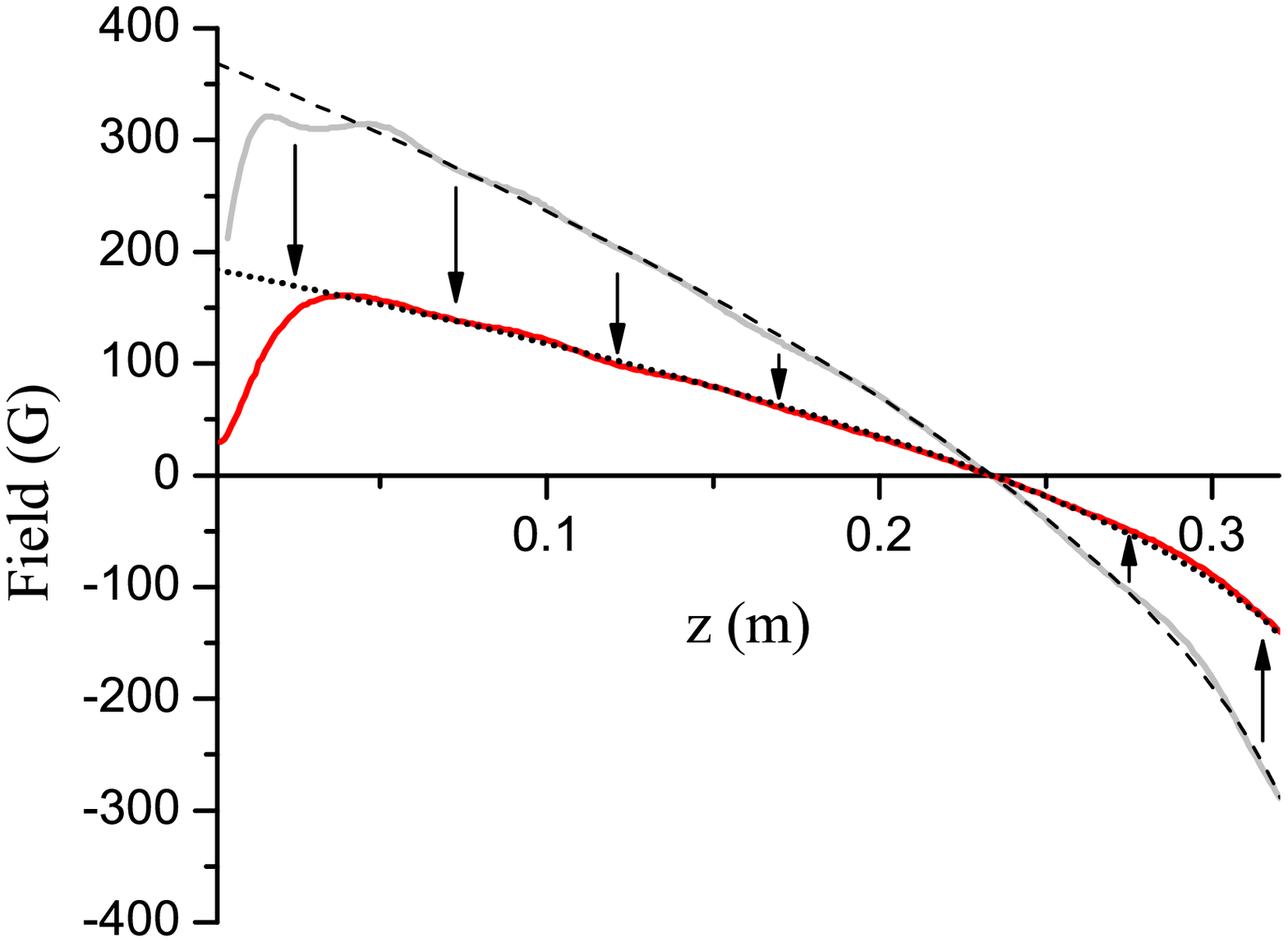}
\label{fig:ProfileDandE}
}
\subfloat[]{
\includegraphics[height=1.65in]{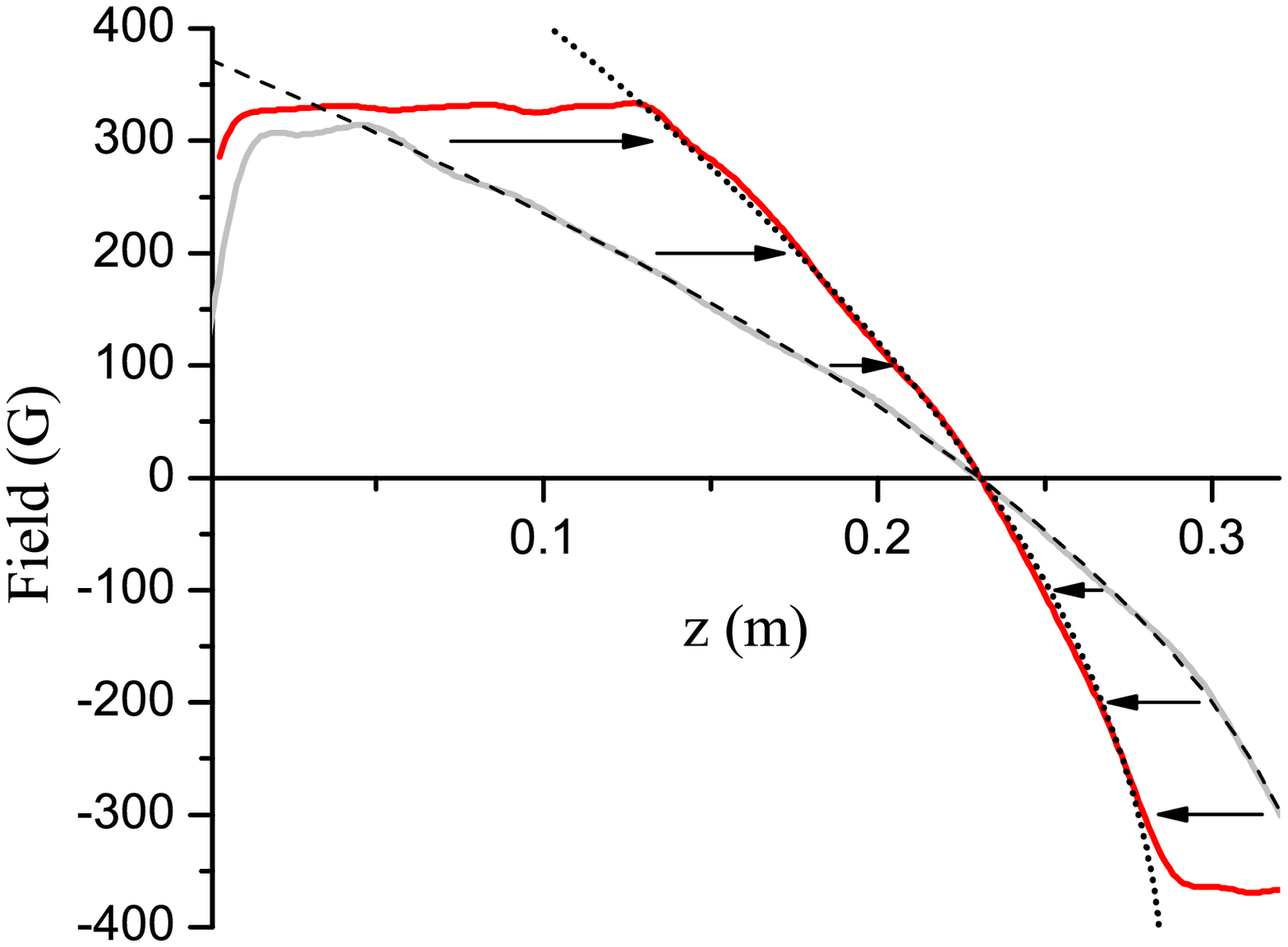}
\label{fig:ProfileKandL}
}
\caption[Optional caption for list of figures]{
(Color online) Magnetic field profiles measured along the axis of the ZS. A typical field profile for the slowing of strontium is represented by the gray line. The dashed line corresponds to a fit using a calculated field profile. The dotted line represents that same profile, but \subref{fig:ProfileDandE} twice weaker at any point of the slower, \subref{fig:ProfileKandL} longitudinally compressed to half-length of the ZS. The red lines are the measured profiles obtained by rearranging the magnets using the servomotors; the switching time is consistently within 0.5 s.
}
\label{fig:ProfilesCompression}
\end{figure}

As a proof of principle, we demonstrate the accuracy and flexibility of the servomotor control by switching the magnetic field between initial \confA, and \confB~giving the same profile but twice weaker at any point of the ZS (Fig.~\ref{fig:ProfileDandE}). We also demonstrate switching from field \confA~to \confC, which corresponds to the same field profile but twice compressed along the propagation axis (Fig.~\ref{fig:ProfileKandL}). Note that the number of magnets on three pairs of rods was changed between Fig.~\ref{fig:ProfileDandE} and Fig.~\ref{fig:ProfileKandL}.

\subsection{Repeatability}
\label{sec:servomotors-repeat}
The repeatability of our servo-controlled ZS was tested by measuring the field profile after switching from different servo configurations to \confA. The field profile can be consistently reproduced within an accuracy of 1\%, reaching the 1 G resolution limit of our gaussmeter at all points in the ZS. Note that to achieve this accuracy, the backlash (mechanical hysteresis) of the servomotor assembly has to be accounted for. This is easily accomplished by systematically instructing the computer to move all the servomotors to the outermost position before adopting the desired configuration.

\section{Real-time optimization of atom slowing}
\label{sec:optimization}
One of the advantages of a motorized ZS is that it can be easily controlled by a computer program. This lends itself well to performing complex optimization algorithms that seek out solutions at a pace unmatched by manual tuning. In the scheme presented here, we feedback on a live measurement of the number of trapped atoms in the MOT.

\subsection{Feedback on the atom loading rate}
\label{sec:optimization-feedback}
A CMOS camera connected via USB to the computer provides a live view of the MOT fluorescence light. The experiment control program gives us the ability to actuate the servomotors, to process images from the camera, and to use a versatile optimizing toolkit \cite{LabExe}.
During an optimization run, the trapping laser beams and the magnetic field of the MOT are turned on continuously, but the repumper lasers are kept off at all times for the following reasons: i) the smaller MOT makes the procedure less sensitive to typical instabilities associated with large atom numbers; ii) the lifetime of the MOT is short under these conditions, and as a result, switching to a new servomotor configuration leads to a stable fluorescence signal on time scales~$\ll 1$~s; iii) due to the artificially large loss rate, we can assume that the loading rate is well represented by the atom number and thus by the fluorescence signal.

Images are gathered from the camera at a rate of~$\sim10$/s. The capture rate of the MOT is inferred by summing the pixel values on the appropriate area of each image, and is then fed into the optimizer which in turn controls the servomotor positions. The detuning of the slowing laser is not a factor in the optimization since  the field profile can be shaped with greater freedom than in traditional slowers, achieving a variety of slopes and maximum field values (Fig.~\ref{fig:ProfilesCompression}). Typically, the optimizer sets a new configuration every second, allowing the servos to fully reach their designated positions and avoiding any transitory effects in the loading or decay of the MOT.

\subsection{Initial servo-configuration}
\label{sec:optimization-initial}
For the study of the automated optimization, we consider the field \confA~from Sec.~\ref{sec:servomotors}. This field closely matches the theoretical profile for strontium (Sr) in the 32~cm designated slowing region (assuming 65\% of maximum deceleration), and overshoots in the region immediately following. Such overshooting is undesirable in an actual slower since it can cause over-slowing, preventing atoms from reaching the trapping region. However, it is illuminating to start the optimization from this field, since it will be shown that our optimization algorithms generate different and equally interesting solutions to this initial condition.

\subsection{Optimization with different algorithms}
\label{sec:optimization-algo}
We have tested several algorithms for the optimization of the MOT capture rate. Since there is no expression for the derivative of the capture rate as a function of the servomotor positions, only black-box optimization algorithms can be used. The results given by two algorithms will be presented here. These two methods are a local search using the Nelder-Mead simplex algorithm, and a global search using a genetic algorithm \cite{NLoptEO}. Genetic algorithms mimic the process of natural evolution by using the concepts of inheritance, mutation and fitness. They perform particularly well on black-box problems with many interdependent variables.
The parameters used for the genetic algorithm runs were typically as follows: population size of 20-40, mutation rate of 40\%, crossover rate of 30\%, and total number of generations of 100-200.

\begin{figure}[ht]
	\centering
		\includegraphics[width=4.5in]{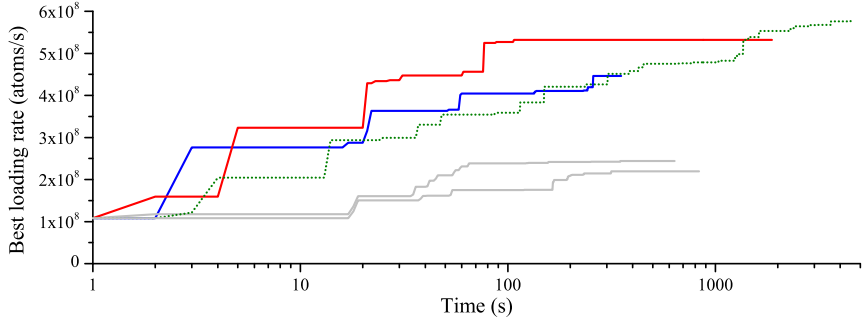}
	\caption{
(Color online) Representation of the best capture rate during typical optimization runs as a function of runtime. The vertical axis scale is labeled using an extrapolated atom loading rate in the presence of repumpers, which is a more conventional measure. Every second the routine probes a new configuration of the 16 servomotor pair positions. The gray lines correspond to runs of a local search \textit{Nelder-Mead simplex} algorithm. The colored lines correspond to a global search using a \textit{genetic} algorithm. The dotted line depicts a genetic optimization run by allowing the 32 servomotors to move independently, i.e. by removing pair symmetry about the ZS axis.
	}
\label{fig:OptimizationRuns}
\end{figure}

Figure~\ref{fig:OptimizationRuns} depicts typical runs of the optimizer. On these plots, the starting point corresponds to the initial \confA, and is used to normalize the atom capture rate for each run, preventing long-term atom number fluctuations from introducing a bias to our comparative analysis. The vertical axis scale is given by the extrapolated capture rate in the presence of repumpers, since this is a more conventional quality measure.

For the local optimization, it takes $\sim100$ servomotor configurations to converge to a magnetic profile that yields an actual loading rate of ~$\sim 2 \times 10^{8}$~atoms/s (as measured with repumpers). The local optimization has to be restarted typically 5-10 times before reaching a steady value. This occurs because it optimizes by probing local configurations that result in small capture rate changes, and hence is sensitive to experimental noise.
Alternatively, the genetic algorithm delivers a loading rate~$\sim 6 \times 10^{8}$~atoms/s in $\sim 1,000$ servomotor configurations. Genetic algorithms are insensitive to noise because they use stochastic operators for selecting the best configurations.  Consecutive runs of the genetic algorithm starting from varying initial conditions produce similar field profiles.  If the 32 servomotors are allowed to move independently without maintaining strict symmetry about the ZS axis, better results can be achieved through compensating any stray fields or imperfections in magnet alignment (Fig.~\ref{fig:OptimizationRuns}), although with a slower initial improvement due to the increased complexity.

\subsection{Interpretation of the optimization results}
\label{sec:optimization-solutions}
\begin{figure}[ht]
	\centering
		\includegraphics[width=4.5in]{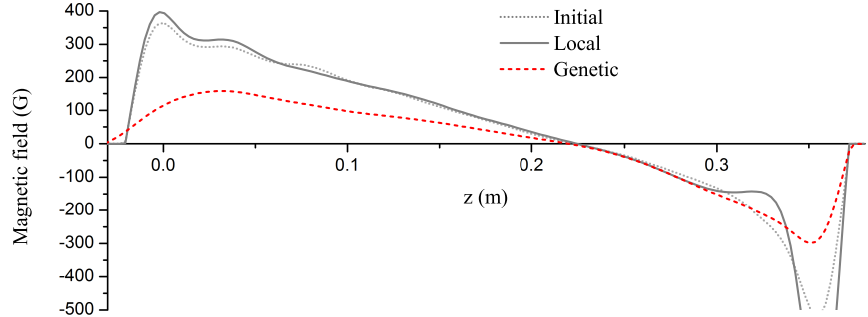}
	\caption{
(Color online) Calculated magnetic field profiles along the ZS for the initial field, locally optimized field, and genetically optimized field. The fields were calculated using final magnet positions and modeling the slower with a corresponding array of magnetic dipoles.
}
\label{fig:ProfileLandG}
\end{figure}

Here we present simulations that give an interpretation of the gains achieved in the two final fields resulting from optimizations starting from the initial field.
Figure~\ref{fig:ProfileLandG} shows the initial field and the two final fields as computationally inferred from the magnet positions.
There are noticeable differences between these three fields: the locally optimized field exhibits an unexpected plateau followed by a steeper overshoot and the genetically optimized field is shallower overall.

These differences can be explained by using a phase portrait, which depicts the deceleration dynamics in the one-dimensional phase-space $(z, v)$ according to the radiation pressure force
$$
F(v,z)=-\,\frac{\hbar \, k \, \Gamma}{2}\frac{s_0(z)}{1 + s_0(z) + 4 \, (\delta_0 + k \, v(z) - \mu \, B(z) / \hbar)^2 / \Gamma^2},
$$
where $B(z)$ is the ZS magnetic field, the average saturation parameter for the cycling transition is chosen to be $s_0(z) = 1.5$, and the detuning of the slowing light is $ \delta_0 = -2 \, \pi \times 505 $ MHz in this case. For Sr, the atomic spontaneous decay rate is $\Gamma=2\pi\times32$ MHz and the optical wave number is $k=2\pi/(461\;\rm{nm})$.

\begin{figure}[h!]
\centering
\subfloat[]{
\includegraphics[width=4.5in]{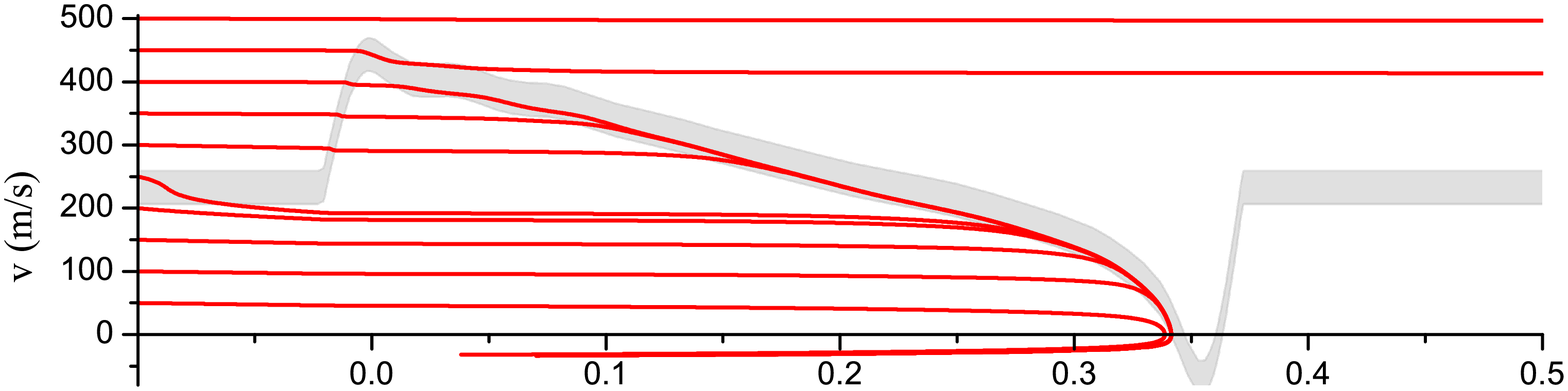}
\label{fig:PhasePortraitsA}
}
\\
\subfloat[]{
\includegraphics[width=4.5in]{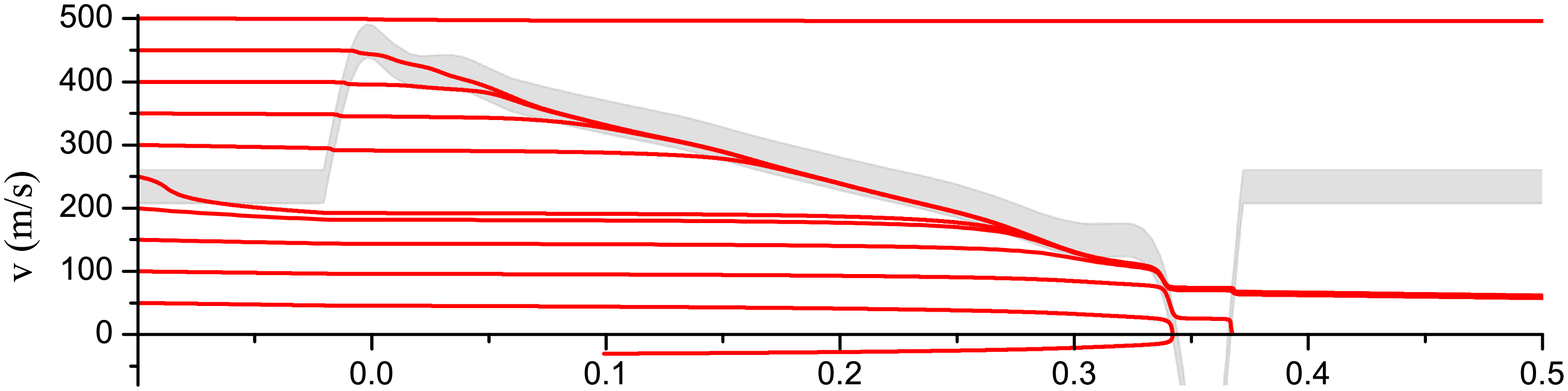}
\label{fig:PhasePortraitsL}
}
\\
\subfloat[]{
\includegraphics[width=4.5in]{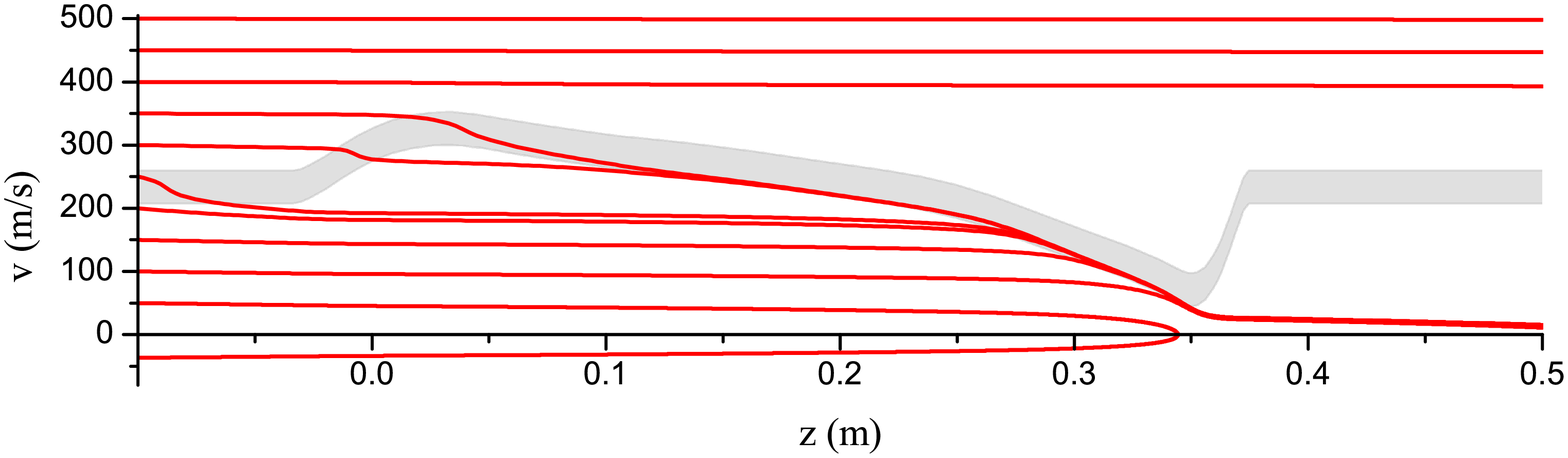}
\label{fig:PhasePortraitsG}
}
\caption[]{
(Color online) Representation of typical atomic trajectories in one-dimensional phase space $(z,v)$ given a magnetic profile $B(z)$ corresponding to \subref{fig:PhasePortraitsA}~the initial profile (\confA), \subref{fig:PhasePortraitsL}~the profile obtained after running a local optimization routine starting from the initial profile, \subref{fig:PhasePortraitsG}~the profile optimized through the use of a genetic algorithm.
The gray zones correspond to the region in which the atoms experience at least a tenth of the maximal radiation pressure force $\hbar\,k\,\Gamma/2$. The atoms enter the ZS at $z\simeq 0.0$ m and the position of the MOT is at $z\simeq 0.5$ m.
}
\label{fig:PhasePortraits}
\end{figure}

Figure~\ref{fig:PhasePortraitsA} shows that the initial profile is over-slowing many of the atoms and reversing their velocities, as described in Sec.~\ref{sec:optimization-initial}.
Figure~\ref{fig:PhasePortraitsL} explains the slowing dynamics of the locally optimized field. The presence of the plateau in the magnetic profile causes the atoms to be slowed less efficiently over the 5 cm region just before the field overshoot. These atoms are then moving fast enough to pass through the region where they were previously stopped, since the interaction time is too short to exert a full stopping force. They do undergo a sudden slowing kick that brings them within the capture range of the MOT.

The genetic algorithm solution depicted in Fig.~\ref{fig:PhasePortraitsG} smoothly reduces the maximum field at the end of the slower ($z\simeq0.3$ m), leading to a velocity distribution that is adequately within the capture range of the MOT, without over-slowing. Note that the convergence to a lower field magnitude near the entrance of the slower, and thus to a weaker deceleration in the first part of the slower, led us to suspect that the typical velocity of the atomic beam is below the 510 m/s design speed. The velocity distribution was subsequently measured, and it was confirmed that the most probable velocity of the Sr beam is only 420 m/s.

\section{Conclusion}
\label{sec:conclusion}
We demonstrated a design and implementation of a servomotor-actuated atomic Zeeman slower that can switch rapidly between different field profiles (including adjusting the slowing region length).  The immediate applications include cooling and trapping of multiple atomic species for quantum optics and precision measurements, as well as real-time optimization of the trapped atom number that allows to adapt to specific experimental circumstances.  The dynamic ZS is remotely controlled with no power consumption in steady state, no water cooling, and a compact design, all particularly suitable for possible future space applications. This approach can be widely used with many ultracold atom and molecule studies, and is a step toward more flexible, cost effective, and automated experiments that are less limited by hardware design choices.

\end{document}